\documentclass[twocolumn]{aastex61}
\usepackage{amssymb}
\usepackage{amsmath}
\usepackage{relsize}		
\usepackage{nicefrac}	

\newcommand{\txd}{{\text{d}}}
\newcommand{\txe}{{\text{e}}}
\newcommand{\calX}{{\mathcal{X}}}
\renewcommand{\leq}{\leqslant}

\newcommand{\Bas}{Basic}
\newcommand{\Spl}{Split}
\newcommand{\Str}{Stretch}
\newcommand{\SpS}{Split\,\&\,Stretch}

\submitjournal{ApJ}

\shorttitle{Failure of Monte Carlo Radiative Transfer}
\shortauthors{Camps \& Baes}

\begin{document}

\title{The Failure of Monte Carlo Radiative Transfer at Medium to High Optical Depths}

\correspondingauthor{Peter Camps}
\email{peter.camps@ugent.be}
\author[0000-0002-4479-4119]{Peter Camps}
\affiliation{Sterrenkundig Observatorium, Universiteit Gent, Krijgslaan 281, B-9000 Gent, Belgium}
\author[0000-0002-3930-2757]{Maarten Baes}
\affiliation{Sterrenkundig Observatorium, Universiteit Gent, Krijgslaan 281, B-9000 Gent, Belgium}



\begin{abstract}
Computer simulations of photon transport through an absorbing and/or scattering medium form an important research tool in astrophysics. Nearly all software codes performing such simulations for three-dimensional geometries employ the Monte Carlo radiative transfer method, including various forms of biasing to accelerate the calculations. Because of the probabilistic nature of the Monte Carlo technique, the outputs are inherently noisy, but it is often assumed that the average values provide the physically correct result. We show that this assumption is not always justified. Specifically, we study the intensity of radiation penetrating an infinite, uniform slab of material that absorbs and scatters the radiation with equal probability. The basic Monte Carlo radiative transfer method, without any biasing mechanisms, starts to break down for transverse optical depths $\tau\gtrsim 20$ because so few of the simulated photon packets reach the other side of the slab. When including biasing techniques such as absorption/scattering splitting and path length stretching, the simulated photon packets do reach the other side of the slab but the biased weights do not necessarily add up to the correct solution. While the noise levels seem to be acceptable, the average values sometimes severely underestimate the correct solution. Detecting these anomalies requires the judicious application of statistical tests, similar to those used in the field of nuclear particle transport, possibly in combination with convergence tests employing consecutively larger numbers of photon packets. In any case, for transverse optical depths $\tau\gtrsim 75$ the Monte Carlo methods used in our study fail to solve the one-dimensional slab problem, implying the need for approximations such as a modified random walk.
\end{abstract}

\keywords{dust, extinction -- methods: numerical -- opacity -- radiative transfer}




\section{Introduction} 

Astronomical observations are shaped by the transport of electromagnetic radiation within and in front of the object being studied. The affecting medium can be, for example, the diffuse interstellar medium, clumpy dust structures near active galactic nuclei, proto-planetary dust disks, or stellar or planetary atmospheres. To help understand the structure and inner workings of the observed systems, researchers develop software codes that simulate the radiation transport in carefully constructed computer models of these objects. Typical models include photon transport in various wavelength ranges, and may focus on emission and absorption lines of the elements in the gas phase and/or on the continuum effects caused by solid dust grains.

The Monte Carlo (MC) radiative transfer (RT) simulation technique is very popular \citep[see, e.g.,][]{Whitney2011,Steinacker2013} because it is conceptually simple, and it supports complex three-dimensional (3D) geometries, multiple anisotropic scattering events, and additional physics such as polarization in a straightforward manner. In the basic MCRT method \citep{Cashwell1959}, the program follows a large but finite number of individual photon packets, each representing a fraction of the total source luminosity, along their journey through the transfer medium. The program draws pseudo-random numbers from the appropriate probability distributions to determine a packet's fate at each event in its life cycle. A packet is terminated when it is absorbed by the transfer medium or when it escapes the system (and possibly is detected by a synthetic instrument). Modern MCRT codes incorporate various variance reduction techniques that involve replacing the physically appropriate probability distribution by a \emph{biased} probability distribution that favors the desired range of outcomes \citep[see, e.g.,][]{Yusef1984, Lucy1999, Niccolini2003, Baes2016}. The resulting bias is compensated by adjusting the \emph{weight} of each photon packet, effectively scaling the luminosity it represents.

The output quantities of a MCRT simulation, such as spatially integrated fluxes or the surface densities in a pixel image, are obtained by summing the contributions of a finite number of photon packets. By the nature of the randomized Monte Carlo process, these quantities are subject to noise, which can generally be reduced by tracing a larger number of photon packets \citep[see, e.g.,][]{Gordon2001}. Individual image pixels naturally show higher noise levels than integrated fluxes because fewer packets contribute to each pixel. Still, in our experience with galaxy modeling \citep[see, e.g.,][]{Baes2010, Delooze2012, Delooze2014, Camps2016}, even noisy images will usually provide a good ``average'' approximation of the correct result. This can be understood intuitively by noting that the MCRT algorithm, by construction, statistically preserves the total luminosity of the source (after subtracting the fraction absorbed by the transfer medium). As a result, we were quite puzzled by the atypical RT simulation results described in the next paragraph.

We recently participated in a benchmark effort by \citet{Gordon2017} comparing the solutions produced by seven RT simulation codes for an externally illuminated slab of dust with well-defined material properties and 3D geometry. For transverse optical depths of the order of unity, the flux penetrating the slab was reproduced by all codes to within less than 10\%. For optical depths $\tau\approx30$, the solutions differed by up to 100\%, and for $\tau\approx75$ they differed by several orders of magnitude. Optical depths of this magnitude are observed in various astrophysical systems, including planetary disks \citep{Jin2016} and active galactic nuclei \citep{Tristram2007}, and they occur in other fields as well, for example shielding in nuclear reactors and for  medical imaging applications. It therefore seems relevant to understand these discrepancies between the various RT codes.

In unpublished tests for optical depths  $\tau\approx75$ with our own 3D dust MCRT code SKIRT \citep{Baes2011,Camps2015}, both the intensity profile along the slab and the integrated flux changed substantially with the number of photon packets launched into the slab, even if the noise level in the intensity profile seemed acceptable. This result implied that the noise level is not a sufficient indicator of convergence when facing higher optical depths in a MCRT simulation, and it prompted us to further investigate the matter and report our findings in the present paper.

In \autoref{Methods.sec} we present a simplified, plane-parallel and single-wavelength version of the \citet{Gordon2017} slab problem (\autoref{Slab.sec}). The one-dimensional geometry allows us to solve the RT problem using spherical harmonics \citep[see, e.g.,][]{Roberge1983, diBartolomeo1995, Baes2001} and using a finite difference approach \citep{Milkey1975, Mihalas1978, Bruzual1988}, providing unambiguous and deterministic reference solutions (\autoref{NumericalMethods.sec}). We also describe our implementation of the MCRT photon life cycle for the plane-parallel slab problem, including two biasing techniques that can be turned on or off (\autoref{MonteCarloMethod.sec}). 

In \autoref{Results.sec} we compare the solutions produced by our MCRT program for varying optical depth and number of photon packets to the corresponding reference solutions, and we show that the behavior noted for the \citet{Gordon2017} benchmark is reproduced in our simplified version. We study the performance of the two optional biasing techniques implemented in our program, and we attempt to understand why the MCRT method breaks down for higher optical depths (\autoref{Failure.sec}). Moreover, we discuss some statistical tests that were developed in the context of MC simulations of particle transport in nuclear physics \citep[see, e.g.,][]{Pederson1997} to help determine whether a particular simulation result is sufficiently converged (\autoref{Statistics.sec}). We then indicate some possibilities for tackling problems that the traditional MCRT methods cannot handle (\autoref{Addressing.sec}). Finally, in \autoref{Conclusions.sec} we summarize our findings.


\section{Methods}
\label{Methods.sec}

\subsection{The plane-parallel slab}
\label{Slab.sec}

\begin{figure}
\centering
\includegraphics[width=\columnwidth]{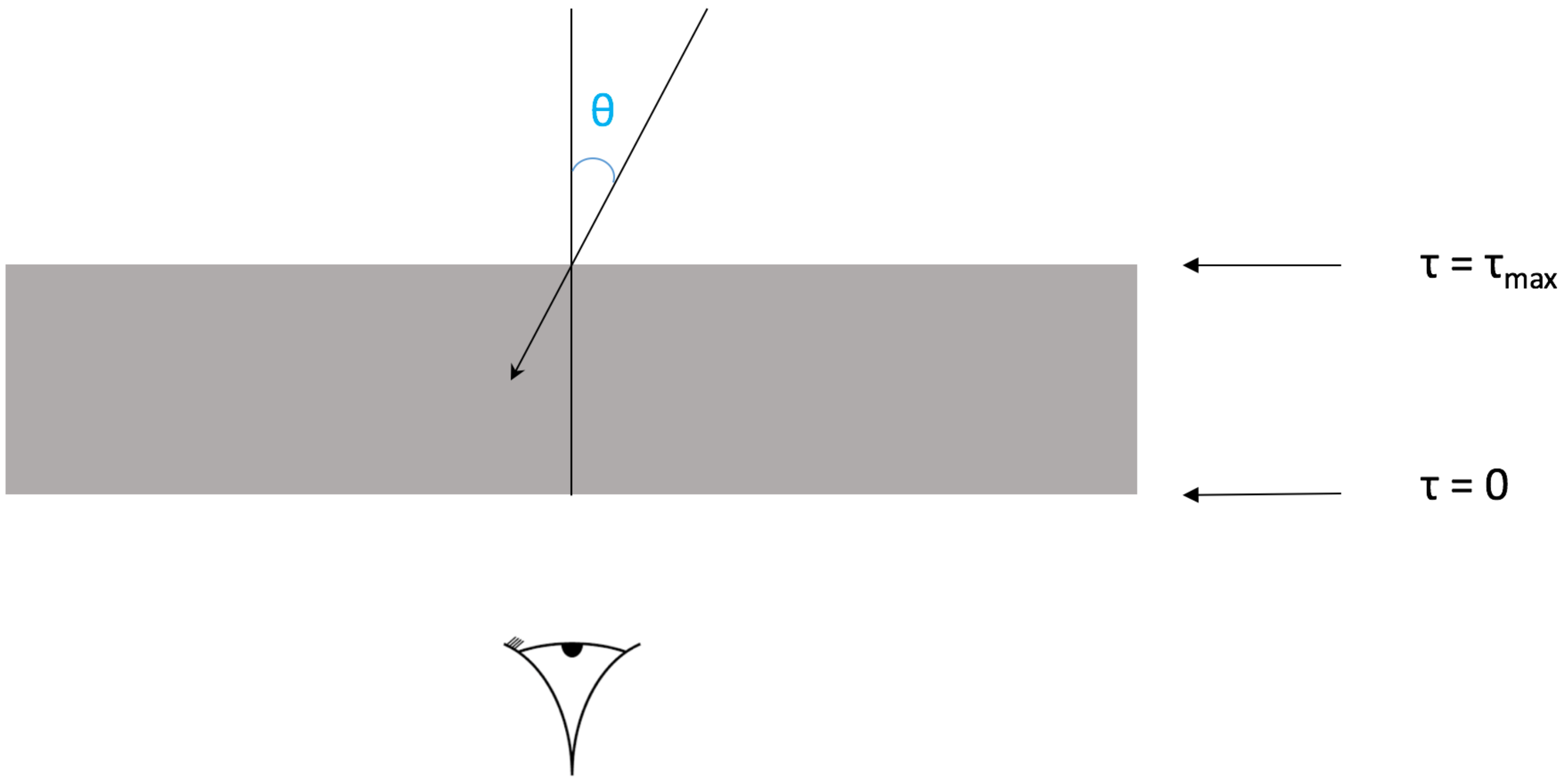}
\caption{Geometry of a plane-parallel slab. The slab is illuminated from the top side, corresponding to $\tau=\tau_{\text{max}}$, by an isotropic radiation field. The observer is located below the slab.}
\label{SlabGeometry.fig}
\end{figure}

Our goal is to solve the RT problem in a plane-parallel slab of uniform density and without internal sources, illuminated from the top side by an isotropic radiation field at a given (arbitrary) wavelength. Instead of the regular cartesian coordinates, we use the optical depth coordinate $\tau$ to indicate the depth into the slab perpendicular to the surface of the slab, so that $\tau=0$ corresponds to the bottom layer of the slab, and $\tau=\tau_\text{max}$ corresponds to the top layer (see \autoref{SlabGeometry.fig}). The angular dependence of the radiation field is parameterized by the cosine $\mu=\cos\theta$ of the angle between the propagation direction and the normal direction. Given these conventions, the time-independent transfer equation for the radiation intensity $I(\tau,\mu)$ can be written as \citep{Baes2001},
\begin{equation}
	\mu\,\frac{\partial I}{\partial\tau}(\tau,\mu)
	=
	I(\tau,\mu)
	-\frac{1}{2}\,\omega
	\int_{-1}^1I(\tau,\mu')\,\Psi(\mu,\mu')\,\txd\mu'
	\label{slabRTE.eq}
\end{equation}
with $\omega$ the scattering albedo and $\Psi(\mu,\mu')$ the angular redistribution function normalized to unity over both incoming and outgoing directions. 
The boundary conditions for the RT problem are
\begin{equation}
	\begin{cases}
	\,\,I(0,-\mu) = 0 \\
	\,\,I(\tau_{\text{max}},\mu) = 1
	\end{cases}
	\qquad\quad\text{for $\mu>0$}.
	\label{slabBounds.eq}
\end{equation}
For the purposes of this paper, we assume that absorption and scattering in the slab material are equally probable and that scattering is isotropic, so that
\begin{equation}
	\omega = \frac{1}{2}
	\qquad\text{and}\qquad
	\Psi(\mu,\mu')=1.
\end{equation}

\subsection{Non-probabilistic numerical methods}
\label{NumericalMethods.sec}

Because of its one-dimensional nature, the plane-parallel slab problem formulated in the previous section can be solved using non-probabilistic numerical methods, including the spherical harmonics method and the finite-difference method discussed below. These methods cannot be readily applied to arbitrary three-dimensional configurations, which is why other methods, such as MCRT, are used in modern codes.

\begin{figure}
\centering
\includegraphics[width=\columnwidth]{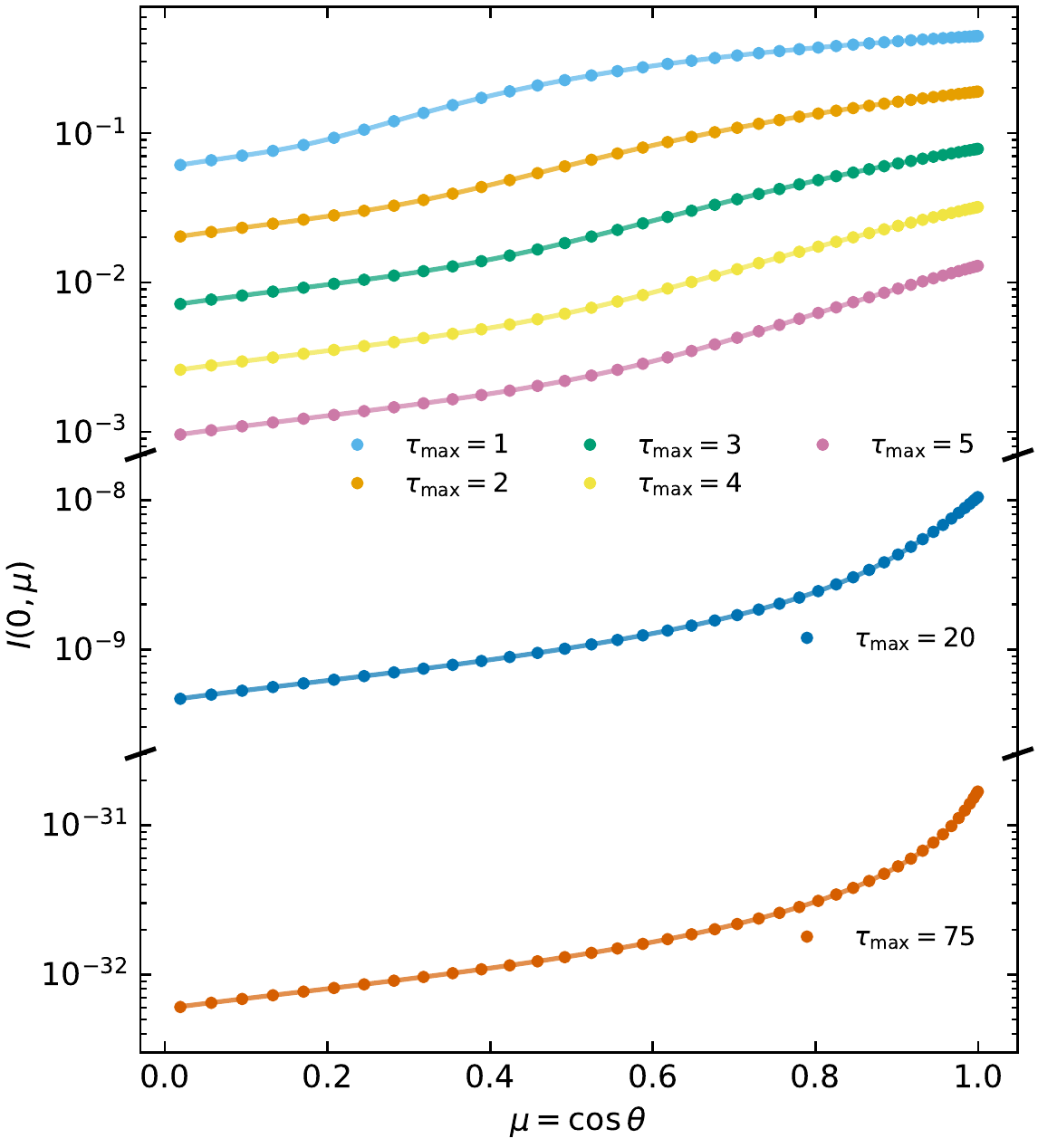}
\caption{Reference solutions calculated with the spherical harmonics method (dots) and the finite-difference method (lines) for the radiative transfer problem illustrated in \autoref{SlabGeometry.fig}, with transverse optical depths from 1 to 75. The vertical axis is ``broken" in two places to allow for the wide intensity range.}
\label{ReferenceSolutions.fig}
\end{figure}

The spherical harmonics method for solving the RT equation was pioneered by \citet{Davison1958}, and later adopted by several others \citep{vandeHulst1970, Flannery1980, Roberge1983, diBartolomeo1995, Corradi1996}. We follow the approach used by \citet{Baes2001} which is based on \citet{Roberge1983} and the adaptations of \citet{diBartolomeo1995}. These authors describe the procedure in detail and we provide just a very brief summary here.

The terms in the RT equation (\autoref{slabRTE.eq}) are expanded in spherical harmonics, which, in plane-parallel symmetry, reduce to the Legendre polynomials $P_L$. Truncating the series expansion at some selected order $L$ leads to an finite set of linear first-order differential equations in the coefficients of the expansion. This set of equations can be reformulated as a classic eigenvalue problem and thus solved by matrix diagonalization. The boundary conditions listed in \autoref{slabBounds.eq} can only be satisfied for $(L+1)/2$ directions $\mu_j$ given by the positive zeros of the Legendre polynomial of order $L+1$. The intensity at an arbitrary direction $\mu$ can be obtained by interpolation.

It turns out that the solution converges even for short series expansions \citep[$L\gtrsim15$ according to][]{Baes2001}. We use a larger value of $L$ for calculating our reference solutions (the dots in \autoref{ReferenceSolutions.fig}) so that we obtain a fair number of intensity values without the need for interpolation. Even with $L=81$ (resulting in intensity values for 41 directions) the calculations complete in a fraction of a second on a modern desktop computer.

The finite-difference method for solving the plane-parallel RT problem is often called Feautrier's method \citep{Feautrier1964}. After its initial formulation, the technique was further extended by authors including \citet{Hummer1971,Milkey1975, Mihalas1978, Bruzual1988} and \citet{Baes2001}, whose approach we follow. In short, the technique relies on the discretization of a differential form of \autoref{slabRTE.eq} on a uniform mesh of $K$ points in optical depth space. The integrals over $\mu$ are approximated by $M$-point Gauss-Legendre quadrature. This procedure eventually leads to a set of coupled equations which can be solved recursively, taking into account the boundary conditions given in \autoref{slabBounds.eq}.

The value of $M$ sets the resolution of the solution in $\mu$ space and it is similar to the value of $(L+1)/2$ in the spherical harmonics method. The second discretization parameter $K$ sets the resolution of the optical depth grid on which the calculations are performed. Its value affects the accuracy of the results, especially at high optical depths. The reference solutions shown as solid lines in \autoref{ReferenceSolutions.fig} were obtained using $K=5000$. This high value was needed to ensure that the finite-difference solution for $\tau_\text{max}=75$ would match the corresponding spherical harmonics solution. Although the finite-difference method is more resource-intensive than the spherical harmonics method, the calculations plotted in \autoref{ReferenceSolutions.fig} still complete in a few seconds on a modern desktop computer.

For a transverse optical depth of $\tau_\text{max}=75$, the intensity penetrating the slab is more than 30 orders of magnitude below the source intensity (bottom curve of \autoref{ReferenceSolutions.fig}). In most astrophysical objects, the faint radiation penetrating such an optically thick barrier would be dominated by radiation from another source located on that side of the barrier, or radiation at a different wavelength would dominate the physics. In those cases, obtaining an accurate simulation result for the penetrating radiation would be noncritical. Still, there might be configurations where even a faint nonlocal radiation component, perhaps Doppler-shifted as a result of relative velocities, would excite a specific atomic or molecular line that would not be triggered by a local radiation field that is much stronger but includes different wavelengths.

\subsection{The Monte Carlo method}
\label{MonteCarloMethod.sec}

The MCRT method essentially follows the individual path of a very large number of photon packets through the transfer medium. The life cycle of each packet is governed by a number of quantities such as the free path length between two interactions, the nature of the interaction (scattering or absorption) and the direction change during a scattering event. Each of these quantities can be described by a random variable, taken from a particular probability distribution. These principles are described in detail by, e.g., \citet{Cashwell1959, Mattila1970, Yusef1984}, and \citet{Bianchi1996}. The MCRT method has been implemented, with various optimizations and refinements \citep[e.g.,][]{Lucy1999, Niccolini2003, Baes2016}, in quite a number of codes in several astrophysical domains \citep[see, e.g.,][for a review of 3D dust continuum RT]{Steinacker2013}.

For the purposes of this paper, we wrote a small computer program that solves the plane-parallel slab problem using four variations of the MCRT method, which we call \Bas, \Spl, \Str, and \SpS, respectively. We now describe these variations in some detail, starting with the \Bas\ method, which uses a bare-bones MCRT life cycle without any biasing.

In the plane-parallel slab geometry (see \autoref{SlabGeometry.fig}), a photon packet is characterized by two variables: the optical depth coordinate $\tau$ (corresponding to a vertical position in the slab) and the direction cosine $\mu$. We launch $N$ identical packets at the top of the slab, i.e.\ with $\tau=\tau_\text{max}$. To obtain an isotropic intensity distribution, we need to take into account the dependence of the luminosity $L$ on the propagation direction given by $\txd L\propto\mu\,\txd\mu\,\txd I$. As a result the normalized probability distribution from which we must sample the direction of packets carrying equal amounts of luminosity becomes, 
\begin{equation}
p(\mu)\,\txd\mu = 2\mu\,\txd\mu, \qquad 0\leq\mu\leq1.
\end{equation}
We can pick a random direction from this distribution using the inversion method,
\begin{equation}
\calX_1 = \int_0^\mu 2\mu'\,\txd\mu' = \mu^2   \quad\Longrightarrow\quad   \mu = \sqrt{\calX_1}
\end{equation}
where $\calX_1$ is a uniform deviate (a random number uniformly distributed over the unit interval).

Next we generate a random free path length $t$ (also in optical depth units) by setting \citep{Baes2001},
\begin{equation}
\calX_2 = \int_0^t \txe^{-\tau}\,\txd\tau = - \txe^{-t}   \quad\Longrightarrow\quad   t = -\ln(1-\calX_2)
\end{equation}
where $\calX_2$ is another uniform deviate. This randomly determined free path length is then compared to the maximum free path length $T$ of the packet under consideration,
\begin{equation}
	T
	=
	\begin{cases}
	\,\,\tau/\mu & \quad\text{if $\mu>0$} \\
	\,\,(\tau-\tau_\text{max})/\mu & \quad\text{if $\mu<0$}.
	\end{cases}
\end{equation}
If $t>T$, the photon packet leaves the slab, and if it is moving downward ($\mu>0$), its direction $\mu$ is recorded. If $t<T$, the packet interacts with the slab material. The nature of this interaction is determined by generating another uniform deviate $\calX_3$ and selecting scattering if $\calX_3<\omega=\nicefrac{1}{2}$ and absorption otherwise.

If the interaction is an absorption, the photon packet disappears and does not contribute to the output intensity. If the interaction is a scattering, the packet acquires a new position and a new direction. Given the free path length $t$ and the original position $\tau$ and direction $\mu$, the new position is easily obtained as $\tau' = \tau-\mu t$. Because we assume isotropic scattering, the new direction is independent of the previous direction and uniform in $\mu$; it is determined from yet another uniform deviate using $\mu' = 2\calX_4-1$. The procedure is now repeated until the photon packet is either absorbed or leaves the slab.

\begin{figure*}
\centering
\includegraphics[width=\textwidth]{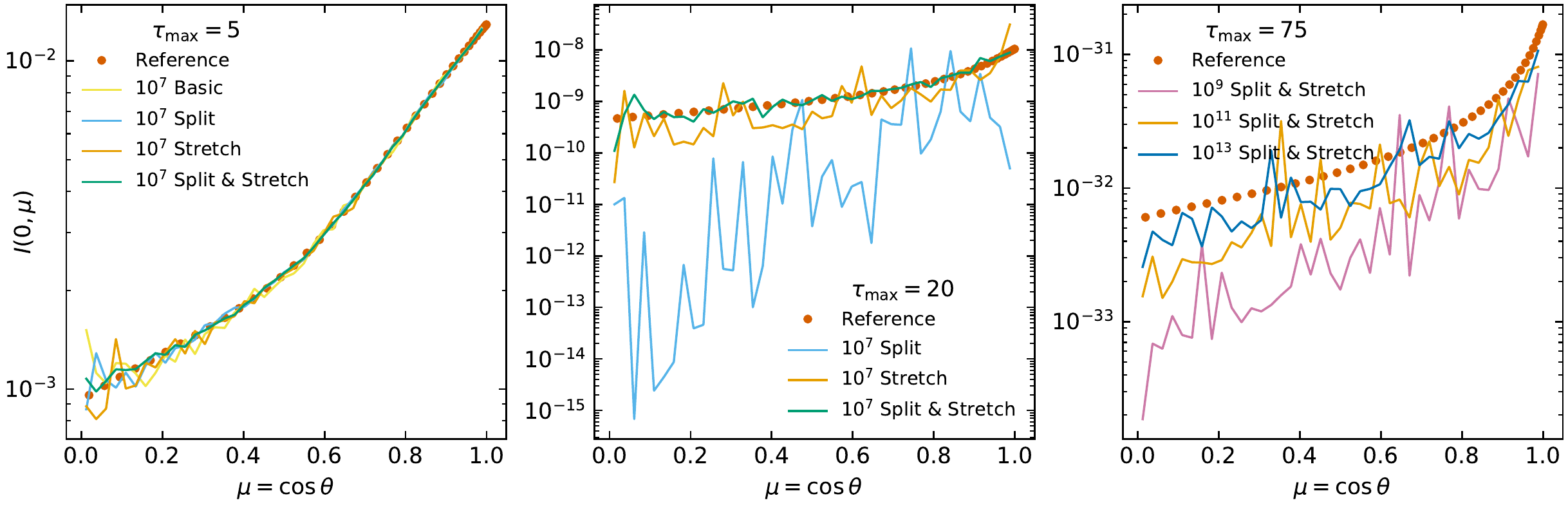}
\caption{Solutions of the slab problem calculated with the Monte Carlo method variations described in the text (solid lines) compared to the corresponding reference solution (red dots) for transverse optical depths 5, 20, and 70 (panels from left to right). The two leftmost panels show results for a fixed number of $10^7$ photon packets, using different methods. The rightmost panel shows results for the \SpS\ method with an increasing number of photon packets.}
\label{PlainSolutions.fig}
\end{figure*}

To capture the angular dependence of the intensity at the bottom of the slab, we setup a grid in $\mu$ space over the interval $[0,1]$ with $M$ equally-sized bins. For each photon packet emerging from the bottom of the slab, we increment the counter for the corresponding $\mu$-bin. When all photon packet life cycles have been completed, we convert these counts to intensities. Using the dependence $\txd I\propto\txd L/(\mu\,\txd\mu)$ yields
\begin{equation}
I_j 
=
\frac{N_j}{N} \, \frac{1/2}{\mu_j} \, \frac{1}{1/M}
=
\frac{N_j}{N} \, \frac{M}{2\mu_j}
\end{equation}
where the index $j$ refers to the $j$th $\mu$-bin. We set $M=41$ for all results in this paper, providing a similar directional resolution as for the reference solutions (but with different spacing because the points in the reference solutions are not equidistant in $\mu$-space).

In the method variations other than \Bas, each photon packet is additionally assigned a weight $w$ that serves as a multiplier adjusting the luminosity carried by the packet. All photon packets are launched at the top of the slab with a ``neutral" weight $w=1$. However, the weights are adjusted along the way as compensation for biasing the probability distributions governing the photon packet life cycle. When detecting a photon packet at the bottom of the slab, we now add its weight to the appropriate bin instead of simply counting it.

In the \Spl\ method, we allow a photon packet to contribute to both absorption and scattering at each interaction site. Instead of randomly choosing the nature of the interaction, the packet is ``split'' in two parts: one that is absorbed and one that scatters and continues its life cycle. In other words, instead of simulating absorption as described above, we always let the packet scatter and we multiply its weight by a bias factor $\nicefrac{1}{2}$ (the value of the scattering albedo $\omega$ under our assumptions), effectively removing the average amount of absorbed luminosity from the packet. Because photon packets are now only terminated when they actually escape the slab, we can expect to properly sample the problem with fewer photon packets. 

In the \Str\ method, we improve the ability of a photon packet to penetrate an optically think medium by biasing the probability distribution for the free path length towards longer or ``stretched" paths. Following the composite biasing technique presented by \citet{Baes2016}, we substitute the physical probability distribution $p(\tau)=\txe^{-\tau}$, which is strongly peaked towards small values, by the biased distribution
\begin{equation}
q(\tau) = \frac{1}{2}\,({\text{e}}^{-\tau} + \alpha\,{\text{e}}^{-\alpha\tau})
\quad\text{with}\;
\alpha=\frac{1}{1+T}
\end{equation}
where the exponential stretch factor $\alpha$ is determined as a function of the current maximum path length $T$ for the photon packet under consideration. To compensate for this subsitution, we multiply the photon packet's weight by the corresponding bias factor,
\begin{equation}
\frac{p(\tau)}{q(\tau)}
= 
\frac{2}{1 + \alpha\,{\text{e}}^{(1-\alpha)\,\tau}}
\leq 2.
\end{equation}
We expect this method to perform better for slabs with higher optical depth.

Finally, the \SpS\ method combines both the \Spl\ and \Str\ biasing mechanisms.


\section{Results and discussion}
\label{Results.sec}

\subsection{The failure of Monte Carlo radiative transfer}
\label{Failure.sec}

Because of the nature of the MCRT method, recalculating a solution with a different (pseudo-)random number sequence yields a different result. However, our tests show that, as expected, multiple runs produce similar solutions, differing solely in the precise shape of the noise pattern. \autoref{PlainSolutions.fig} shows representative solutions of the slab problem presented in \autoref{Slab.sec} calculated with various Monte Carlo methods (\autoref{MonteCarloMethod.sec}, solid lines), compared to the corresponding reference solution (\autoref{NumericalMethods.sec}, red dots). 

The leftmost panel in \autoref{PlainSolutions.fig} presents the results for all four methods and a slab with transverse optical depth $\tau_\text{max}=5$, using a fixed number of $10^7$ photon packets. It is fair to conclude from this figure that the discussed MCRT methods indeed work as advertised. The noise level increases for directions away from the normal, i.e.\ for direction cosines $\mu\lesssim0.3$. Close to the parallel direction ($\mu\lesssim0.1$), the radiation intensity is an order of magnitude lower than close to the normal direction ($\mu\gtrsim0.9$), which means that fewer photon packets contribute to these bins, resulting in higher noise levels. The signal-to-noise ratio can easily be improved by launching more photon packets (not shown in the figure but confirmed by our tests). Closer inspection of the results for $\mu\lesssim0.3$ reveals that the noise levels differ between the methods; for example, the \SpS\ method shows less noise than the other methods. This is not surprising, because the biasing techniques used in this method have been designed to reduce the variance in the calculated result. 

The performance differences between the methods become much more prominent in the middle panel of \autoref{PlainSolutions.fig}, which presents results for $\tau_\text{max}=20$, still using the same number of $10^7$ photon packets. The solution for the \Bas\ method is not shown, because in most tests not a single photon packet actually reaches the bottom of the slab. Indeed, according to the reference solution (red dots in middle panel of \autoref{PlainSolutions.fig}), the observed intensity is more than 8 orders of magnitudes lower than the source intensity. In other words, the probability that one of the $10^7$ photon packets penetrates the slab is smaller than $10^7/10^8=0.1$. The biasing mechanisms included in the other methods improve the situation, with varying success.

As a first step, the absorption-scattering split implemented in the \Spl\ method causes many more photon packets to penetrate the slab, albeit with reduced weights (multiplication by the albedo for each scattering event) and at the expense of spending more time per photon packet (because packets are terminated only when they escape at either side of the slab). The result calculated by the \Spl\ method (blue line in the middle panel of \autoref{PlainSolutions.fig}) is not only very noisy, it is also off by many orders of magnitude, and the running average has the incorrect shape (the deviation from the reference solution is much larger for smaller values of $\mu$). The \Str\ method performs much better (orange line), because the path length stretching technique is specifically designed to penetrate high-optical depth barriers more easily. Because of the longer free paths, packets interact much less frequently with the slab medium, and thus don't get terminated so easily by absorption. The \SpS\ method (green line), combining both techniques, performs best, although it still suffers from noise fluctuations close to an order of magnitude for $\mu\lesssim0.1$.

When raising the transverse optical depth to $\tau_\text{max}=75$, even the \SpS\ method fails to produce a correct solution without using an excessive number of photon packages. The rightmost panel in \autoref{PlainSolutions.fig} shows results for the \SpS\ method with $\tau_\text{max}=75$ and an increasing number of photon packets. While the result seems to converge to the correct solution for an increasing number of packets, even with $10^{13}$ packets (dark blue line) the MCRT result still underestimates the solution for most directions, and it shows noise fluctuations of about an order of magnitude. This calculation, with $10^{13}$ packets for $\tau_\text{max}=75$, consumed 52 days of equivalent serial time on a multi-core computer system. 

\begin{figure}
\centering
\includegraphics[width=\columnwidth]{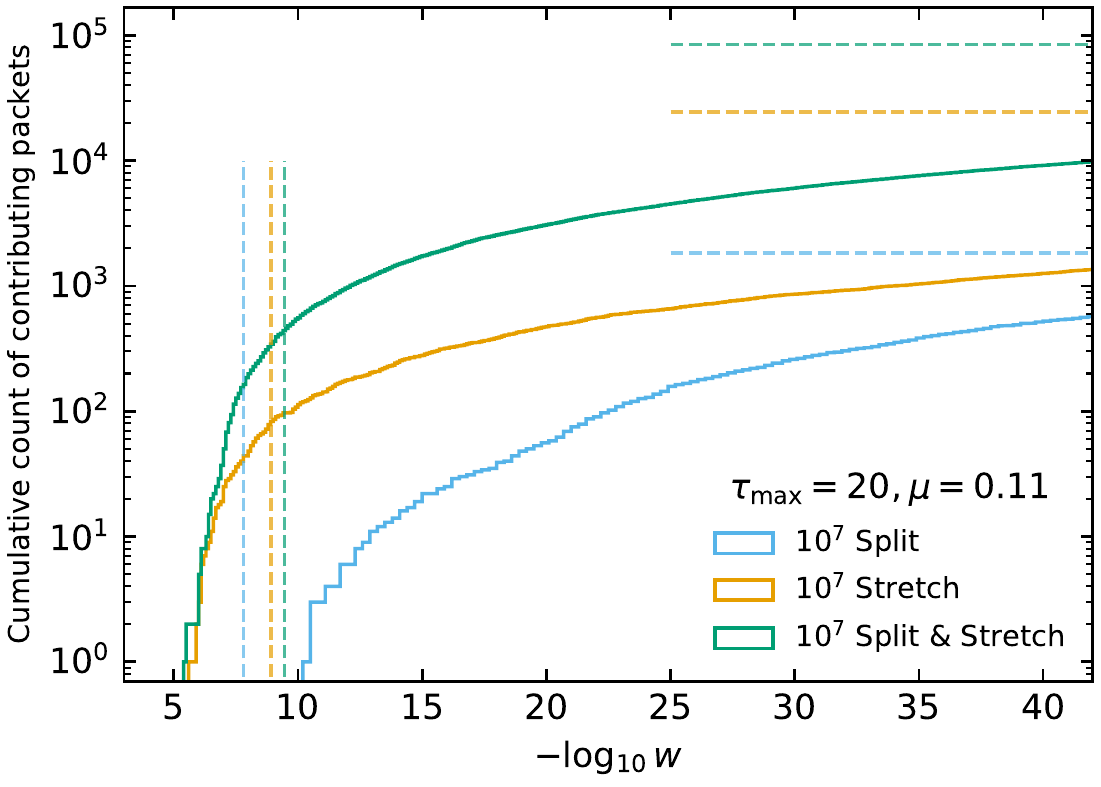}
\caption{Cumulative distribution of the photon packet contributions $w_i$ to the observed intensity in the direction bin centered on $\mu\approx0.11$ for a slab with $\tau_\text{max}=20$ and $N=10^7$, corresponding to the solutions shown in the middle panel of \autoref{PlainSolutions.fig}. The horizontal dashed lines indicate the total number of packets contributing to the bin, or in other words, the limit of each cumulative distribution. The vertical dashed lines are explained in the text.}
\label{WeightDistribution.fig}
\end{figure}

Reviewing the three panels in \autoref{PlainSolutions.fig}, we might generalize (at least for the plane-parallel slab problem) that the ``running average'' of the MCRT solution generally matches the corresponding reference solution as long as the transverse optical depth is sufficiently low ($\tau_\text{max}\lesssim20$) and the MCRT method employs the appropriate variance reduction techniques. For high optical depths ($\tau_\text{max}\gtrsim75$) the MCRT technique becomes intractable. Moreover, when considered in isolation and without further evaluation, any of the unconverged results (e.g. the blue line in the middle panel and the purple and orange lines in the rightmost panel of \autoref{PlainSolutions.fig}) might be mistaken as a proper result, albeit somewhat noisy. To develop ways for avoiding such misinterpretation, we need to understand why MCRT fails for these higher optical depths, and why the average result depends so heavily on the number of photon packets in the simulation.

A first intuitive insight can be gained from considering the distribution of the individual photon packet contributions $w_i$ to the observed intensity in a particular direction bin. To properly handle the extremely small photon packet weights resulting from repeated application of the biasing factors, we perform the MCRT calculations for this part of our study using floating point numbers with 50 significant digits and a sufficiently large exponent range. \autoref{WeightDistribution.fig} shows the cumulative distributions of the contributions for the bin centered on $\mu\approx0.11$ ($\theta\approx84^\circ$) in the solutions of the middle panel of \autoref{PlainSolutions.fig}, i.e.\ for a slab with $\tau_\text{max}=20$ and $N=10^7$. For the \Spl\ method (blue curve), 1835 photon packets reach this particular bin (indicated by the horizontal dashed blue line). Many of the contributions are very small; the smallest one is $w_\text{min}\approx4\times10^{-379}$ (i.e., far outside the range shown in the figure). With an albedo of $\omega=\nicefrac{1}{2}$, this weight corresponds to 1257 consecutive scattering events. The largest contribution is $w_\text{max}\approx6\times10^{-11}$, corresponding to 34 scatterings. The vertical dashed blue line indicates the weight each of the 1835 arriving packets would need to contribute on average to reach the total weight that matches the reference solution for the bin. Because even the largest contribution for the \Spl\ method is several orders of magnitude below this value, it is clear that a much larger number of photon packets would need to be launched for the simulation to approach the correct solution.

The situation for the \Str\ and \SpS\ methods (orange and green in \autoref{WeightDistribution.fig}) is much improved because many more photon packets contribute to the bin, and the largest contributions are much larger than those for the \Spl\ method. The steeper climb of the cumulative distribution for the \SpS\ method implies that it performs better than the \Str\ method, but it is not clear from this qualitative analysis whether the result can be considered to be sufficiently converged.

\subsection{Statistical tests for convergence}
\label{Statistics.sec}

A basic statistical treatment to obtain a confidence interval for astrophysical dust MCRT simulation results is provided by, e.g., \citet{Gordon2001} and \citet{Steinacker2013}. More detailed statistical techniques, however, are routinely employed by nuclear particle transport simulation codes. We specifically refer to the Overview and Theory manual for MCNP \citep[hereafter ``the MCNP manual",][]{MCNP2003}. MCNP is a general Monte Carlo N-Particle transport code developed by Los Alamos National Laboratory in the United States. The procedures described in the MCNP manual are based on early work by \citet{Estes1978} and further research by, e.g., \citet{Dubi1979, Forster1991, Forster1992, Booth1992a, Booth1992b} and \citet{Pederson1997}. \citet{Dunn2012} provide a succinct summary of the key concepts. For our analysis in this section, we largely follow the definitions and recommendations presented in the MCNP manual.

The total MCRT simulation result $W$ for a given observed radiation direction bin (before converting to intensity) is obtained by accumulating the individual photon packet weights $w_i$, i.e.\ $W=\sum w_i$, where the index $i$ and the sum run over the $N$ photon packets launched during the simulation. By the nature of the MC method, the weights $w_i$ are independently sampled\footnote{For this to be true, all observed effects directly or indirectly caused by a particular launched photon packet must be combined into a single contribution.} from some unknown probability distribution function $f(w)$. Note that this distribution may have a discrete component. For example, for the \Bas\ method, the weights are either 0 or 1, and for the \Spl\ method, the weights are multiples of the albedo.

If the expectation value and variance of the unknown distribution $f(w)$ exist and are finite, the central limit theorem applies and we can estimate both quantities from the sample values $w_i$.\footnote{If the transverse optical depth of the slab is high enough relative to the number of photon packets launched, the number of packets detected in a given bin might be so low that the central limit theorem does not apply and the sample mean is not normally distributed. For example, by performing 500 runs using different random seeds, we verified that the sample mean for the $\tau_\text{max}=20$ cases shown in \autoref{WeightDistribution.fig} does not follow a Gaussian distribution, while it does when the optical depth is lowered to, e.g., $\tau_\text{max}=5$ (with the same number of photon packets). However, in case the sample mean is not distributed normally, we expect the sample variance to be sufficiently large that the analysis presented in this section will still trigger the proper alarms. We also verified that this is the case for the examples shown in \autoref{WeightDistribution.fig}.} We define the sample mean $\bar{w}$ and the estimated $k$-th central moment $M_k$ of the distribution as
\begin{align}
\bar{w} &= \frac{1}{N}\,\sum w_i,
\\
M_k &=  \frac{1}{N}\,\sum (w_i-\bar{w})^k.
\label{Mk.eq}
\end{align}
In a MCRT simulation, $N$ is always sufficiently large to assume $N\approx N-1$, so there is no need to correct for small sample sizes. We can then write the estimated variance $S^2_{\bar{w}}$ of the sample mean as
\begin{equation}
S^2_{\bar{w}} = \frac{1}{N}\,M_2.
\end{equation}
We define the relative error $R$ as
\begin{equation}
R \triangleq \frac{S_{\bar{w}}}{\bar{w}} = 
\left[ \frac{\sum w_i^2}{(\sum w_i)^2} - \frac{1}{N} \right]^{\nicefrac{1}{2}},
\label{R.eq}
\end{equation}
where the right-hand side shows the expansion in sums of $w_i$ and $w_i^2$, which can be easily tracked during the simulation.

If all $w_i$ values are nonnegative, as is the case when detecting photon packet luminosities, it is easy to see from its definition that $0 \leq R \leq 1$. If all $w_i$ are zero, the ratio in \autoref{R.eq} becomes undefined, and we arbitrarily set $R=1$ because, for our slab problem, a null solution is certainly incorrect.\footnote{For problems where a null solution is physically possible, one could set $R=0$.} If all $w_i$ are nonzero and equal, $R$ reaches its minimum value of zero. If there is only a single nonzero contribution, $R$ approaches its maximum value of unity. As a final limiting case, consider a result with a small number of $n \ll N$ nonzero and equal contributions, such as when using the \Bas\ method for a slab with medium transverse optical depth. In this case, $R\approx 1/\sqrt{n}$, recovering the behavior of Poisson noise. 

\begin{figure}
\centering
\includegraphics[width=\columnwidth]{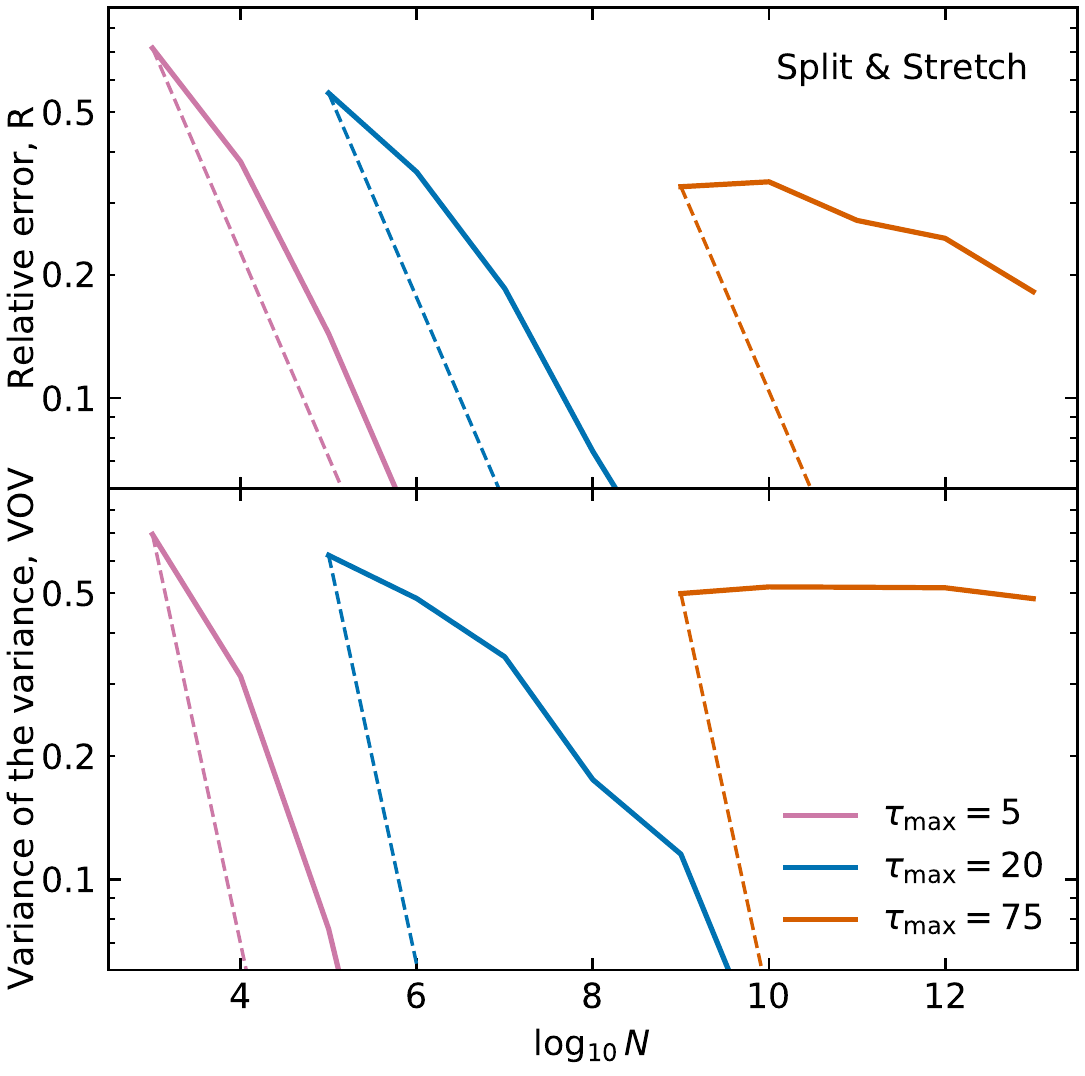}
\caption{Relative error $R$ (top) and variance of the variance VOV (bottom) averaged over all direction bins as a function of the number of photon packets launched to tackle the slab problem using the \SpS\ method, for the transverse optical depths also shown in \autoref{PlainSolutions.fig}. The dashed lines indicate the slope of the expected power-law dependence ($R\propto N^{\nicefrac{-1}{2}}$, $\text{VOV}\propto N^{-1}$).}
\label{Statistics.fig}
\end{figure}

The top panel of \autoref{Statistics.fig} shows the relative error, averaged over all direction bins for each solution, as a function of the number of photon packets launched to solve the plane-parallel slab problem with the \SpS\ method. The solid lines plot the evolution of the average $R$ for the three transverse optical depth values for which selected solutions are shown in \autoref{PlainSolutions.fig}. Based on theoretical considerations and experience with actual simulation results, the MCNP manual recommends that $R$ should be smaller than $0.1$ for the corresponding result to be considered reliable. In the range $0.1<R<0.2$, results are dubbed ``questionable", and for $R>0.2$, results are generally unreliable. For $\tau_\text{max}=5$ the \SpS\ method easily reaches the $R=0.1$ mark at $N\approx10^{5.3}$, and for $\tau_\text{max}=20$ at $N\approx10^{7.6}$. For $\tau_\text{max}=75$, however, the relative error is still at the ``questionable" level even with $N=10^{13}$ photon packets. These assessments are compatible with a visual evaluation of the errors in the solutions shown in \autoref{PlainSolutions.fig}.

It is worth noting that \citet{Gordon2001} define uncertainties similarly to the definition of $R$ in \autoref{R.eq}.\footnote{\citet{Gordon2001} do not aggregate the effects from multiple scattering events of a single photon packet to a single contribution, which may to some extent skew comparison with their results.} Because the models in their study have optical depths of the order of unity, the resulting uncertainties remain well under the $0.1$ ``danger'' value. As a consequence, the \citet{Gordon2001} suggestion to employ the uncertainties as relative error bars is justified. Uncertainties with values above $0.1$, however, should be interpreted as indicating questionable or unreliable results rather than as a relative error bar.

The dashed lines in \autoref{Statistics.fig} (top panel) indicate the slope of the expected evolution of $R$ as $N$ increases (with an arbitrary starting point to guide the eye). The dependency can easily be derived from \autoref{R.eq} to be $R\propto 1/\sqrt{N}$. For $\tau_\text{max}=5$ and $20$ the actual evolution of $R$ approaches the expected power-law behavior, while for $\tau_\text{max}=75$ the slope differs significantly even at $N=10^{13}$. This is another clear indication that the solution has not yet converged and thus should be considered unreliable.

Continuing to follow the MCNP manual and the prior research it is based upon, we further define the variance of the variance, VOV, a quantity that measures the relative statistical uncertainty in the estimated $R$. The definition assumes that the third and fourth central moments of the probability distribution $f(w)$ exist and are finite. The VOV is much more sensitive to large fluctuations in the $w_i$ values than is $R$, and it can thus detect situations where the obtained $R$ value is unreliable. The VOV can be written in terms of the moments given by \autoref{Mk.eq} as 
\begin{equation}
\text{VOV} \triangleq \frac{1}{N}\,\frac{M_4-M_2^2}{M_2^2}
= \frac{\sum(w_i-\bar{w})^4}{\sum(w_i-\bar{w})^2}-\frac{1}{N}.
\label{VOV.eq}
\end{equation}
The quantity can also be expressed as a function of the sums $\sum w_i^k$, $k=1,2,3,4$, which can be tracked during the simulation.

Based on statistical experiments, the MCNP manual recommends that the VOV should be below 0.1 to ensure a reliable confidence interval.\footnote{The value 0.1 is convenient, which is why the VOV is used rather than its square root.} It can be easily seen from \autoref{VOV.eq} that the VOV is expected to decrease as $1/N$. The bottom panel of \autoref{Statistics.fig} shows the VOV, averaged over all direction bins, as a function of $N$, for the same solutions as those considered before. The dashed lines again indicate the expected evolution for sufficiently large $N$. For $\tau_\text{max}=5$ the VOV behaves as expected even for small $N$, resulting in convergence criteria that are more relaxed than those for $R$ (top panel). For higher optical depths, however, the VOV decreases more slowly than expected. For $\tau_\text{max}=75$, the VOV barely decreases even up to $N=10^{13}$, adding another reason for caution with regard to the reliability of the solution.

Our analysis shows that the statistical quantities $R$ and VOV can be successfully employed to evaluate the reliability of a MCRT solution (without a priori knowledge of the correct solution), at least for the plane-parallel slab problem considered in this work. If both $R$ and VOV are below 0.1, and both quantities decrease with $N$ as expected, one can be reasonably confident to have a reliable solution.

\begin{figure}
\centering
\includegraphics[width=\columnwidth]{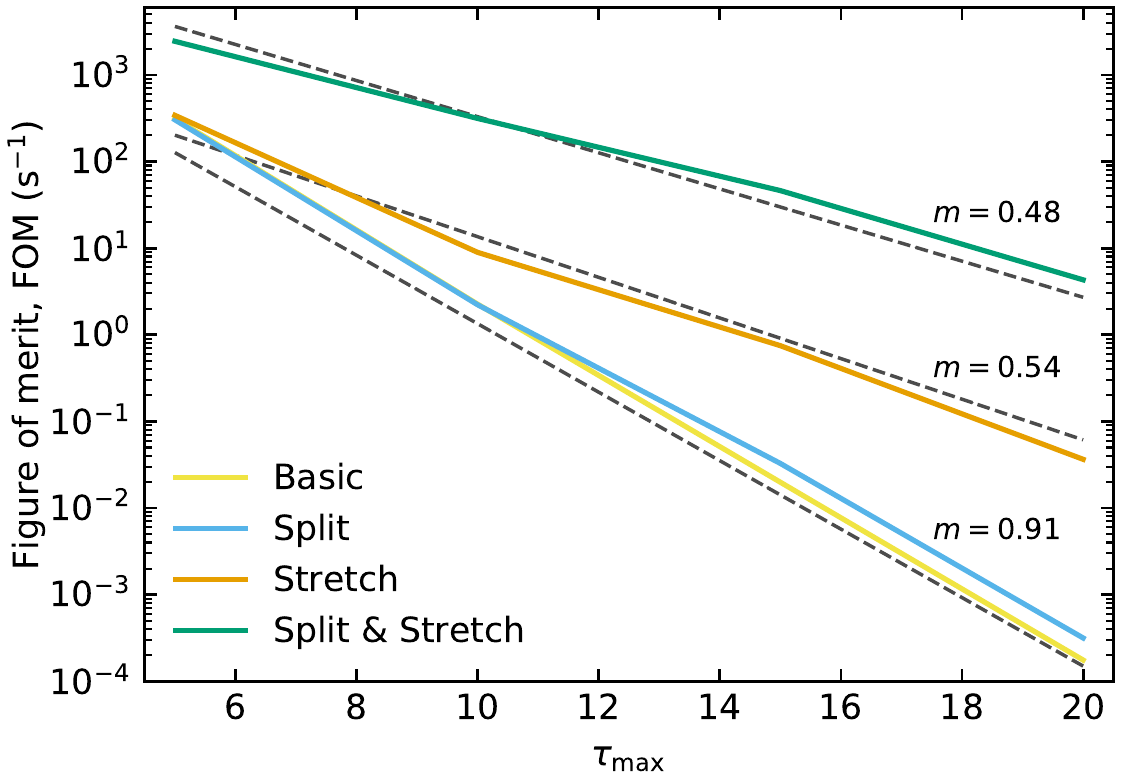}
\caption{Converged figure of merit, $\text{FOM}=1/(R^2T)$, for each of the four MCRT method variations considered in this work as a function of the transverse optical depth $\tau_\text{max}$ of the slab (solid lines). The dashed lines trace exponential functions proportional to $\txe^{-m\tau_\text{max}}$ for three values of the exponent index $m$.}
\label{FigureOfMerit.fig}
\end{figure}

Another useful quantity is the figure of merit, FOM, defined as
\begin{equation}
\text{FOM} = \frac{1}{R^2\,T},
\label{FOM.eq}
\end{equation}
where $T$ is the computer time spent on the simulation in seconds. Because the computer time is proportional to the number of photon packets $N$ and the square of the relative error should scale as $1/N$, the FOM should be approximately constant once the solution has converged. The evolution of the FOM for increasing $N$ can thus be used as a reliability indicator; it is easier to verify constant behavior than to evaluate scaling with a power law. The FOM also provides a measure for the efficiency or ``merit" of a particular method for solving the problem at hand. A higher FOM indicates that the simulation can reach a given level of precision in less computer time.

The solid lines in \autoref{FigureOfMerit.fig} plot the (converged) FOM for each of the four methods considered in this work as a function of the transverse optical depth $\tau_\text{max}$. The value of the FOM depends on the implementation details of the program and scales with the processing speed of the computer system in use. For the data shown in this figure, we ran our C++ MCRT program on a regular desktop computer using a single execution thread. While the details may vary for other computer systems and implementations, the overall trends should be representative. The \Bas\ and \Spl\ methods show similar performance. While the \Bas\ method requires for a much larger number of photon packets, the \Spl\ method calculates many more path segments for each photon packet (because the packets are forced to continue after each interaction). Apparently the efficiency of both mechanism is in balance. The \Str\ method performs significantly better, which is understandable because it has been designed to penetrate high-optical-depth barriers. Somewhat unexpectedly, adding absorption/scattering splitting to path length stretching (as in the \SpS\ method) substantially boosts performance again. We postulate that the number of additional path segments introduced by absorption/scattering splitting is now limited because the stretched path segments are much longer, on average.

The dashed lines in \autoref{FigureOfMerit.fig} trace exponential functions of the form $A\,\txe^{-m\,\tau_\text{max}}$ for three values of the exponent index $m$ and with arbitrary proportionality constant $A$, chosen to approximately match the respective solid lines. It is clear from the figure that, as $\tau_\text{max}$ increases, the FOM for each method decreases exponentially. Equivalently, the processing time needed to solve the problem with given precision rises exponentially as $T\propto\txe^{m\,\tau_\text{max}}$. This can be expected because of the exponential relation between the penetrating intensity and the optical depth of the barrier. Reading the appropriate data point from \autoref{FigureOfMerit.fig}, and using \autoref{FOM.eq}, we can deduce that the \SpS\ method requires about 25~s to reach a precision of $R=0.1$ at $\tau_\text{max}=20$. Using the exponent index of $m=0.48$ for this method, we can extrapolate the time to reach a similar precision at $\tau_\text{max}=75$ to be more than $7\times10^{12}$~s or $2\times10^{5}$~yr. On a high-performance computer with 512 nodes of 16 cores each, the calculation would still require more than 27 years. In other words, the problem is not tractable on present-day computers with the MCRT methods considered in this work.

\subsection{Addressing the problem}
\label{Addressing.sec}

It is somewhat ironic that, for the purposes of this work, we had the MCRT method consume many days of computer time to obtain a solution that can be calculated in a fraction of a second by a non-probabilistic numerical method. However, as indicated in \autoref{NumericalMethods.sec}, for arbitrary three-dimensional problem configurations this ``fast lane" is no longer available. So the question becomes how to augment or complement the MCRT method so that it can handle higher optical depths.

As a first step, it is crucial to equip MCRT codes with the statistical instruments discussed in the previous section, so that users can evaluate the accuracy of the simulation results, and decide whether the reported inaccuracy is relevant for the physics of the broader problem. Conversely, the statistics can be used to determine, manually or automatically, when there is a need for enabling specific, possibly approximative, high-optical-depth techniques.

To this end, several authors have introduced schemes in which the transport equation is replaced by a diffusion equation in regions of the spatial domain where the diffusion approximation can be justified. Indeed, \citet{Larsen1987} show that both equations are asymptotically equivalent when three key assumptions hold: spatial variations of physical quantities are slow, the physical medium is optically thick, and interactions with the physical medium are scattering-dominated. For a discretized spatial domain, the first two assumptions translate to requiring that the radiation field and the optical properties of the medium can be considered constant within each spatial cell, while at the same time each cell still contains many free path lengths. In these hybrid schemes, the MCRT mechanism is used to solve the transport equation in non-diffusive regions, and another, more efficient, method is used to solve the diffusion equation in diffusive regions. For example, the discrete diffusion Monte Carlo method \citep[DDMC, e.g.,][]{Densmore2007, Densmore2012, Abdikamalov2012, Cleveland2015} allows radiation packets to step from one diffusive spatial cell to another following the solution of a  diffusion equation discretized on those cells, as opposed to performing a possibly long-winded random walk within each cell.

An alternate technique for diffusive regions, usually referred to as modified random walk (MRW), was introduced by \citet{Fleck1984} and further refined and generalized by \citet{Min2009}, \citet{Robitaille2010}, and \citet{Keady2017}. The MRW procedure picks a sphere centered on the current radiation packet's position that encloses a region of space with constant medium properties  (e.g., within a single spatial cell). Given the assumptions for the diffusive regime, this sphere may still enclose a large number of MC random-walk path segments. To replace this long random walk by a single step, the procedure estimates the probability function for the total length of the path enclosed in the sphere, which is then used to determine the resulting extinction, or equivalently, the amount of energy absorbed by the transfer medium. The original MRW procedure assumes isotropic scattering \citep{Fleck1984}, and corrections can be made for anisotropic scattering \citep{Min2009}.

A number of caveats remain to be considered. To our knowledge, the above techniques have been implemented mostly (or solely) in combination with an immediate re-emission scheme assuming energy balance under local thermal equilibrium conditions. The smaller dust grains and the hydrocarbon compounds in the interstellar medium, however, are often not in equilibrium with the radiation field \citep{Sellgren1984, Boulanger1988, Helou2000} and are characterized by a temperature probability distribution rather than a single equilibrium temperature. The emission spectrum of these grains depends on the surrounding radiation field in a highly nonlinear way \citep{Guhathakurta1989, Draine2001, Camps2015b}. Modern dust MCRT codes thus need to keep track of the absorbed energy at each frequency (in each spatial cell), so that the diffusive regime solution mechanism must consider individual frequencies (or narrow ranges) rather than integrating over a broad range of frequencies.

Most importantly, at ultraviolet and optical wavelengths, a typical interstellar dust mixture has a scattering albedo between 0.2 and 0.65 \citep{Draine2003, Zubko2004, Jones2013}. In other words, scattering interactions do not dominate absorption events. Because one of the key assumptions of the diffusion methods does not hold, they are, in principle, not applicable to interstellar dust. The approximation may be acceptable in certain parameter regimes (albedo, scattering anisotropy, optical depth), but this cannot be assumed without further consideration. Perhaps some of the diffusion methods can be adjusted to handle a broader parameter range, for example by adopting probability density functions in the MRW technique that take into account albedo and scattering anisotropy parameter values in a broader range.

The plane-parallel slab configuration presented in \autoref{Slab.sec} would form an excellent test-bed in this context. The non-probabilistic numerical methods described in \autoref{NumericalMethods.sec} are easily extended to support arbitrary albedo and anisotropic scattering, so that reference solutions would be readily available for comparison with the results of adding a diffusive regime procedure to the implementation of the Monte Carlo method described in \autoref{MonteCarloMethod.sec}. Such a study could pave the way for implementations of these methods in fully equipped 3D dust MCRT codes, although this may require additional effort. While some of the hybrid MC/diffusion techniques have been implemented in 3D codes \citep[e.g.,][assuming equilibrium heating]{Robitaille2011}, most studies were performed with axisymmetric or even spherically symmetric models \citep[e.g.,][]{Min2009, Pinte2009, Densmore2007, Densmore2012, Abdikamalov2012, Cleveland2015}. Adaptations that handle non-equilibrium dust grain heating and a wider albedo and scattering anisotropy parameter range should be properly tested with high-optical depth 3D model configurations similar to the \cite{Gordon2017} benchmark.

At the same time, an improved understanding of which methods can properly solve the 1D slab problem for various high-optical depth parameter regimes, and why this is the case, might lead to more fundamental insights. For example, while the statistical tests discussed in \autoref{Statistics.sec} can help detect the failure of a method after an alleged solution has been obtained, it would be beneficial to develop a theoretical or empirical framework to detect such failure before actually attempting to solve the problem.


\section{Summary and conclusions}
\label{Conclusions.sec}

We set out to investigate the apparent failure of the Monte Carlo (MC) Radiative Transfer (RT) method noted by \citet{Gordon2017} for optical depths of about $30$ or more. To this end, we formulated a plane-parallel, single-wavelength RT problem consisting of an infinite slab of uniform material with given transverse optical depth $\tau_\text{max}$. We summarized two non-probabilistic methods that can be used to generate reference solutions for this one-dimensional configuration. We also described our implementation of four variations of the Monte Carlo method to solve the same problem, including the \Bas\ photon packet life cycle and two optional biasing techniques called \Spl\ (absorption-scattering split) and \Str\ (path length stretching). 

We compared the solutions produced by our MCRT program to the corresponding reference solutions in \autoref{PlainSolutions.fig}. The photon packets in the \Bas\ method essentially fail to penetrate the slab for $\tau_\text{max}$ as low as $20$. The \Str\ biasing technique has a substantial positive effect, as expected, because it was designed to improve penetration of higher optical depths. The best results are obtained with the \SpS\ method, which combines both biasing techniques. However, a slab with $\tau_\text{max}=75$ forms an insurmountable barrier even for this optimized photon packet life cycle. A calculation with $10^{13}$ photon packets, consuming 52 days of equivalent serial computer time, still does not recover the correct solution because too few photon packets with a significant contribution penetrate the slab.

We subsequently discussed and applied a number of statistical tests originally developed by other authors for evaluating MC simulations of nuclear particle transport. The relative error, $R$, is defined as the ratio of the estimated standard deviation over the estimated mean of the individual photon packet contributions to a particular detector bin. The variance of the variance, VOV, defined similarly, serves in turn as a reliability measure for $R$. To calculate these statistics, the simulation must keep track of the sums $\sum w_i^k$, $k=2,3,4$, in addition to the regular accumulated contribution $\sum w_i$. We verified that the recommendations developed for the MCNP nuclear particle transport code \citep{MCNP2003} are also valid for our slab problem (see \autoref{Statistics.fig}). In short, when both $R$ and VOV are below 0.1 and show the expected statistical dependency with an increasing number of photon packets launched, the corresponding result can be considered to be reliable. 

We also discussed a quantity called figure of merit, defined as $\text{FOM}=1/(R^2T)$, where $R$ is the relative error and $T$ is the amount of computer time spent on the simulation (see \autoref{FigureOfMerit.fig}). We used the FOM to determine that calculating a solution for the $\tau_\text{max}=75$ slab problem would take dozens of years even on a powerful high-performance computer system.

We conclude that the MCRT methods used in our study start to fail for barriers with transverse optical depths as low as $20$, and that detecting the anomalous results requires the judicious application of statistical tests, possibly in combination with convergence tests employing consecutively larger numbers of photon packets. While we focused our tests on a plane-parallel geometry and fixed material properties, changing the geometry and including materials with anisotropic scattering or a different scattering albedo will most likely leave the overall conclusions intact. The benchmark results reported by \citet{Gordon2017} at least point into that direction.

The above implies the need for approximative methods to handle high optical depths in MCRT simulations. An excellent candidate is the modified random walk technique, which was already implemented and tested by other authors in specific circumstances. Because reference solutions can easily be obtained for the one-dimensional slab problem presented in this work, we suggest that it can play a significant role in testing the accuracy of the modified random walk method for various optical properties of the transfer medium.


\acknowledgments

\section*{Acknowledgments}

We thank the co-authors of the TRUST slab benchmark paper \citep{Gordon2017}, and specifically the lead author, Karl Gordon, for many interesting discussions that inspired us to investigate further and write the current paper.

\bibliography{FailedMCRT}




\end{document}